\documentclass[11pt]{amsart}
\usepackage{geometry}
\usepackage{graphicx}
\usepackage{amssymb, amsmath, mathabx, amsthm, stmaryrd, braket}
\usepackage{epstopdf}
\usepackage{natbib}
\usepackage{xcolor}

\DeclareMathOperator{\Ima}{Im}

\title{Review of a Quantum Algorithm for Betti Numbers}
\author{Sam Gunn and Niels Kornerup}
\address{The University of Texas at Austin}
\email{samgunn111@utexas.edu, nielskornerup@utexas.edu}
\date{June 17, 2019}

\begin{document}

\begin{abstract}
We looked into the algorithm for calculating Betti numbers presented by \cite{LGZ} (LGZ). We present a new algorithm in the same spirit as LGZ with the intent of clarifying quantum algorithms for computing Betti numbers. Our algorithm is simpler and slightly more efficient than that presented by LGZ. We present a thorough analysis of our algorithm, pointing out reasons that both our algorithm and that presented by LGZ do not run in polynomial time for most inputs. However, the algorithms could run in polynomial time for calculating an approximation of the Betti number to polynomial multiplicative error when applied to some class of graphs for which the Betti number is exponentially large, if a good lower bound on the eigenvalues of a particular matrix can be found.
\end{abstract}

\maketitle

\section{Background and Definitions} \label{section:background}
The $k$th Betti number $\beta_k$ is essentially the number of $k$-dimensional holes in a topological space.\footnote{We use the term ``$k$-simplex/Betti number/etc" to refer to what is normally a $(k-1)$-simplex/Betti number/etc, because it simplifies the presentation of the algorithm.} The algorithm presented by LGZ aims to calculate the Betti numbers of simplicial complexes. The input to the algorithm is a graph, although the motivation lies in the application to graphs constructed in a particular way. In this section we will briefly describe the required background for the algorithm, starting from a distance matrix for points representing a topology we are interested in. We assume the reader knows what a simplex is, but only a very rudimentary understanding is necessary.

Given an $n \times n$ distance matrix $D$ and a parameter $\epsilon > 0$, let $G$ be the $n \times n$ adjacency matrix for the graph where vertices are adjacent if they are at most a distance of $\epsilon$ away from each other in $D$. For positive integer $k$, we can define a simplicial complex on the $n$ vertices where for $h \le k$, the $h$-simplices are the $h$-cliques of $G$. This is called the Vietoris-Rips complex.

\begin{figure}
    \centering
    \includegraphics[width=6cm]{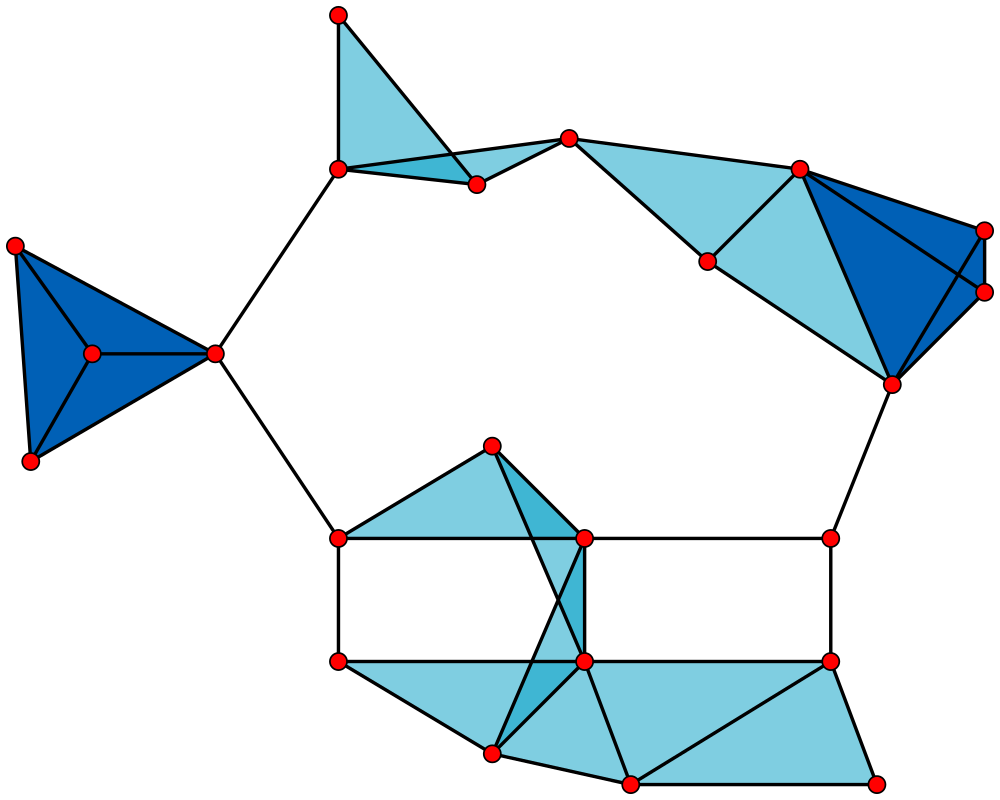}
    \caption{Example Vietoris-Rips complex. Points within $\epsilon$ of each other have a line drawn between them.}
    \label{fig:vietorisrips}
\end{figure}

Let $S_k$ be the set of $k$-simplices in the Vietoris-Rips complex, and write $s \in S_k$ as $[j_1 \hdots j_k]$, where $j_i$ is the $i$th vertex in $s$. Let $\mathcal{H}_k$ be the abstract complex vector space with basis $S_k$. Elements of $\mathcal{H}_k$ are called chains. We define the boundary map $\partial_k : \mathcal{H}_k \to \mathcal{H}_{k-1}$ by its action on the basis $S_k$:
$$\partial_k([j_1 \hdots j_k]) = \sum_{i = 1}^k (-1)^{i-1} [j_1 \hdots \hat{j_i} \hdots j_k]$$
where $\hat{j_i}$ indicates that $j_i$ is not included in the list. We extend $\partial_k$ linearly to $\mathcal{H}_k$.

\begin{figure}
    \centering
    \includegraphics[width=10cm]{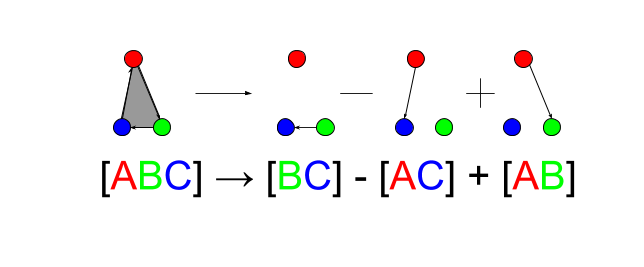}
    \caption{Example application of the boundary map $\partial_3$.}
    \label{fig:boundary}
\end{figure}

The quotient space
$$\ker \partial_k / \Ima \partial_{k+1}$$
is called the homology, and its dimension is the $k$th Betti number $\beta_k$. Elements of the homology are cycles (boundary-less chains), where two cycles are considered equal if there is a continuous deformation of one to the other within the simplicial complex (i.e., if their difference is a boundary).

It turns out that the (combinatorial) Laplacian $\Delta_k = \partial_k^\dagger \partial_k + \partial_{k+1} \partial_{k+1}^\dagger$ satisfies
$$\ker \Delta_k \cong \ker \partial_k / \Ima \partial_{k+1}$$
If we take $\partial_1 = 0$, then $\Delta_1$ is the familiar graph Laplacian to which the matrix-tree theorem applies. An analogous interpretation holds for higher dimensions (\cite{MR}).

\section{Overview of the Algorithm}
Given an $n$-vertex graph $G$ representing a Vietoris-Rips complex, and an integer $k$ specifying the dimension of interest, our algorithm calculates $\beta_k$. At a high level, it proceeds in two stages: First, we prepare the state $\rho_k$, a uniform mixture over $\mathcal{H}_k$; then we perform phase estimation on $\Delta_k$ with $\rho_k$ as input to estimate $\beta_k = \dim (\ker \Delta_k)$.

\section{State Preparation} \label{section:prep}
Let $S_k$ be the set of $k$-simplices in $G$. We will represent $s \in S_k$ with vertex set $\{v_1 \ldots v_k\}$ as a string of length $n$ with Hamming weight $k$, where there are ones at indices $v_1 \ldots v_k$ and zeroes elsewhere. For the phase estimation step, it will be convenient later to have the mixed state $$\rho_k = \frac{1}{|S_k|} \sum_{s \in S_k} \ket{s}\bra{s}$$

To make $\rho_k$ we will first make the state $$\ket{\psi_k} = \frac{1}{\sqrt{|S_k|}}\sum_{s \in S_k} \ket{s}$$
This state can be prepared using an unknown-number-of-target variant of Grover's algorithm as can be found in \cite{BBHT}, where marked items are taken to be simplices (as can be checked in $k^2$ gates).

We can also implement the transformation $P_k$ such that $P_k\ket{0} = \ket{\psi_k}$ and $P_k$ is a unitary if we initially apply approximate counting to estimate $|S_k|$ (as in \cite{BHTA}), and then implement $P_k$ using Grover's algorithm with the fixed number $|S_k|$ of marked items. We will use $P_k$ in Section \ref{section:sampling}.

In order to avoid searching over all $2^n$ strings of length $n$, we can encode the Hamming weight $k$ strings as natural numbers. This requires a one-time cost of $\tilde{O}(kn^2)$ gates and $\tilde{O}(n k)$ additional gates per round of Grover's algorithm and is described in detail in Section \ref{section:combinatorial-number-system}. Using this encoding, we can create the state $\ket{\psi_k}$ using $$\tilde{O}\left(kn^2 + nk \sqrt{ \binom{n}{k} /|S_k|}\right)$$
gates. Once we have $\ket{\psi_k}$, we can apply a CNOT gate to each qubit in $\ket{\psi_k}$ into ancilla zero qubits to yield $\rho_k$.

\section{Phase Estimation} \label{section:phaseestimation}
Now we need to calculate the dimension of the kernel of the Laplacian $\Delta_k$. Since $\Delta_k$ is not sparse, we will instead use phase estimation on the $n$-sparse Hermitian matrix
$$B = \begin{bmatrix}
0 & \partial_2 & 0 & 0 & 0 & \hdots \\
\partial_2^\dagger & 0 & \partial_3 & 0 & \hdots \\
0 & \partial_3^\dagger & 0 & \hdots \\
 0 & 0 & \hdots & & & 0 \\
 0 & \hdots & & & 0 & \partial_n \\
 \hdots & & & 0 & \partial_n^\dagger & 0 \\
 \end{bmatrix}$$
which satisfies
$$B^2 = \begin{bmatrix}
\Delta_1 & 0 \\
0 & \Delta_2 & & \hdots \\
& & \hdots \\
& \hdots & & \Delta_n
\end{bmatrix}$$
We need to keep an extra $\log n$ qubits to keep track of the simplex dimension, $k$, which indexes the Laplacians in $B^2$. Since $\ker B = \ker B^2$, we can just initialize these extra qubits to specify $k$ in the computational basis when estimating $\dim (\ker \Delta_k)$. Because $B$ is Hermitian, $n$-sparse, and has entries $-1$, $0$, or $1$, we can implement $U = e^{iB}$ in $\tilde{O}(n^2)$ gates (\cite{BCK}).

At this point we must be careful about our domain, as the restriction of $U$ to simplex dimension $k$ acts on all of $\mathbb{C}^{2^n}$, but we are only interested in $\ker \Delta_k$ in $\mathcal{H}_k \subset \mathbb{C}^{2^n}$. Fortunately this is resolved by simply using $\ket{k} \bra{k} \otimes \rho_k$ as input to phase estimation with $U$.

Suppose we use $\ket{k} \ket{v}$, where $\ket{v} \in \mathcal{H}_k$, as input to the phase estimation algorithm with $U$. Since $\Delta_k$ is invariant on $\mathcal{H}_k$ (\cite{Friedman}), an eigenbasis $E_k$ for $\Delta_k$ will have as a subset an eigenbasis $E_k'$ that spans $\mathcal{H}_k$, so suppose further that $\ket{v}$ is an eigenvector of $\Delta_k$. If $\ket{v} \in \ker \Delta_k$, then it is clear that $B \ket{k} \ket{v} = 0$. If $\ket{v} \not\in \ker \Delta_k$, then $\ket{k} \ket{v}$ will be decomposed into eigenvectors of $B$ with non-zero eigenvalues.\footnote{If $\lambda$ satisfies $\Delta_k \ket{v} = \lambda \ket{v}$, and $\ket{k} \ket{v} = \sum \ket{w_i}$ for eigenvectors $\ket{w_i}$ of $B$, then the associated eigenvalues $\lambda_i$ all square to $\lambda$ and are therefore all non-zero.}

Now if we run the phase estimation algorithm with $U$ on $\ket{k} \bra{k} \otimes \rho_k$, the probability of measuring 0 in the eigenvalue register is exactly $\dim (\ker \Delta_k) / \dim \mathcal{H}_k = \beta_k / |S_k|$, as long as we use enough qubits to distinguish zero from non-zero eigenvalues.

Let $\lambda_{max}$ and $\lambda_{min}$ be the largest and smallest eigenvalues of $\Delta_k$, respectively. Then by scaling down $\Delta_k$ by $1/\lambda_{max}$ to avoid multiples of $2\pi$, it takes $\tilde{O}(\log (\lambda_{max}/\lambda_{min}))$ qubits in the eigenvalue register, and thus $\tilde{O}(\lambda_{max}/\lambda_{min})$ applications of $U_k$, to distinguish eigenvalues from 0. By the Gershgorin circle theorem, $\lambda_{max} \in O(k)$, but we do not know of any reason why $1/\lambda_{min}$ should be polynomial for the general combinatorial Laplacian.

\section{Sampling} \label{section:sampling}
After preparing the state in Section \ref{section:prep}, we have $\sum_{v \in E_k'} \ket{v}\ket{v}$, which is $\rho_k$ in the first register. By Corollary 4 in \cite{BHTA}, we can use quantum counting to exactly compute the multiplicity of the zero eigenvalue for the Laplacian using $\tilde{O}(\sqrt{|S_k|\beta_k})$ rounds in Grover's algorithm. Since we want to apply the Grover diffusion operator over $S_k$ rather than $\{0,1\}^n$, we will need to use $P_k(I-2\ket{0}\bra{0})P_k^\dag$ (where $P_k$ is the state preparation circuit from Section \ref{section:prep}) instead of the standard diffusion operator. Our evaluation function will run the phase estimation algorithm and invert the states $\ket{v} \ket{v} \ket{\lambda_v}$ where $\ket{\lambda_v}=0$. Thus, each round of Grover's requires running both our state preparation and phase estimation algorithms. This brings the gate complexity of our entire algorithm for computing $\beta_k$ to
\begin{equation} \label{eq:1}
    \tilde{O}\left(\sqrt{\beta_k |S_k|} \left[nk \sqrt{\binom{n}{k}/|S_k|} + \frac{n^2k}{\lambda_{min}}\right]\right)
\end{equation}
Clearly, there is no hope of this ever being polynomial unless $\lambda_{min}$ is bounded away from 0 and $k$ is constant -- but there already exist classical algorithms that run in time $\tilde{O}(\binom{n}{k}^2)$.

\section{Remarks}
\subsection{Complexity of LGZ}
We ended up with a different gate complexity than LGZ. They said that the complexity of sampling eigenvalues of $\Delta_k$ to multiplicative error\footnote{Our understanding of the algorithm suggests that this should be additive error, which arises from using $\delta^{-1}$ applications of $U_k$ during phase estimation.} $\delta$ is
$$\tilde{O}\left(n^5 \delta^{-1} \sqrt{\binom{n}{k}/|S_k|}\right)$$
And that of computing $\beta_k$ to multiplicative error $\delta$ is
$$\tilde{O}\left(n^5 \delta^{-1} \sqrt{\binom{n}{k}/\beta_k}\right)$$
The algorithm presented by LGZ multiplied the complexities of state preparation and phase estimation because their $U_k$ included a projection operator, which we do not think is necessary due to the existence of $E_k'$ and our discussion around it (see Section \ref{section:phaseestimation}).

This number appears to bypass certain details that we have considered here, including searching over Hamming weight $k$ strings and bounding the eigenvalues away from 0. Furthermore, it is still only polynomial in $n$ and $k$ if $\beta_k$ is very large and one just wants an estimate to polynomial multiplicative error. If we want $\beta_k$ exactly, we need $\delta \approx 1/\beta_k$ and the complexity closely resembles ours:
$$\tilde{O}\left(n^5 \sqrt{\beta_k \binom{n}{k}}\right)$$
This is only polynomial if $k$ is constant and $\beta_k$ is small. Nonetheless, it is a significant speedup over classical algorithms, if a good bound for $\lambda_{min}$ can be found.

Note that for most practical applications it is probably expected that $\beta_k$ is small, but there are extreme cases. For complexes in $\mathbb{R}^d$, \cite{Goff} shows that $\beta_k \in O(n^k)$, but that for $d$ as small as 5 there are examples of classes of graphs where $\beta_k \in \Omega(n^{k/2+1/2})$. We do not know of examples where $\beta_k \in \Omega(\binom{n}{k} / \text{poly}(n,k))$, as would be necessary to make the multiplicative-error algorithm polynomial.

\subsection{Persistence}
Although calculating Betti numbers could be interesting, persistent Betti numbers are of more interest. With $\epsilon$ as in Section \ref{section:background}, a persistent Betti number is the rank of the map on homology
$$\ker \Delta_k |_{\mathcal{H}_k^\epsilon} \to \ker \Delta_k |_{\mathcal{H}_k^{\epsilon+p}}$$
induced by the inclusion of the corresponding chain complexes $\mathcal{H}_k^\epsilon \to \mathcal{H}_k^{\epsilon+p}$, where $p > 0$. We were not able to extend the algorithm to persistent computations. It is not obvious that this is possible, because the standard algorithm for classically computing persistent homology (as presented by \cite{ZC}) requires extensive element-wise manipulation of matrices, and we do not know of any analogue to the Laplacian that has similarly nice properties for persistent homology.

Note that computing Betti numbers at several levels of the filtration is not sufficient to deduce persistent Betti numbers. For example, imagine a set in $\mathbb{R}^2$ consisting of the vertices of distant squares labeled by $i=0,\hdots,m$, where the $i$th square has edge length $2^{i/2}$. As soon as the $i$th square's diagonal is connected, the $(i+1)$th square becomes a cycle. Then for all $\epsilon \ge 1$, $\beta_2 = 1$, but that has no connection with persistence.

LGZ suggest that one create an equal superposition over $(\epsilon, k)$ pairs and run the entire algorithm in superposition with those parameters, but we don't see how the desired information can be recovered from that.

\bibliographystyle{plainnat}
\bibliography{references}

\section{Appendix}
%

\subsection{Combinatorial number system} \label{section:combinatorial-number-system}
In order to avoid searching over the full $2^n$ sized space of length $n$ strings, we can encode the Hamming weight $k$ strings as natural numbers. The combinatorial number system gives us a bijection between the Hamming weight $k$ strings and the set $\{1, \hdots, \binom{n}{k}\}$ (\cite{BFT}).

Using the combinatorial number system, converting from a Hamming weight $k$ string with 1's at indices $x_1, x_2 \ldots x_n$ where $x_1 < x_2 \ldots < x_n$ to its corresponding natural number is given by $\sum_{i = 1}^n \binom{x_i}{i}$.

Now we need a way to convert from a natural number $l \in \{1, \hdots, \binom{n}{k}\}$ to its corresponding Hamming weight $k$ string. Since the combinatorial number system represents Hamming weight $k$ strings in their lexicographic order, we know that the largest value of $x$ such that $\binom{x}{k} < l$ will be the position of the first 1 in our string. We can then recursively solve the problem with $l' = l - \binom{x}{k}$ and $k' = k - 1$. Once we know the locations of all the ones, writing down our output takes $O(n)$ gates.

We can use Pascal's triangle to create a look-up table for all relevant binomial coefficients with $O(n^2)$ addition operations. Since the largest entry in this table is $\binom{n}{k}$, we get that all entries are at most $\tilde{O}(k)$ bits long, giving a gate complexity of $\tilde{O}(n^2 k)$ to create the table. Finding the largest value of $x$ such that $\binom{x}{k} < l$ can be done via binary search in $\tilde{O}(k)$ gates using the lookup table. 

Since we need to do this $k$ times, the total gate complexity of converting from a natural number to its Hamming weight $k$ string representation is $\tilde{O}(k^2+n)$. Likewise, converting from a Hamming weight $k$ string to the corresponding natural number can be done with the look-up table with $\tilde{O}(kn)$ gates. Since $n\geq k$, the combinatorial number system conversion takes a total of $\tilde{O}(kn)$ gates.

\end{document}